\documentclass[12pt]{article}
\usepackage[pctex32]{graphics}
\begin{document}
\vspace{1cm}
\begin{center}
~\\
{\bf  \Large Thermodynamics on Noncommutative Geometry in Coherent State Formalism}
\vspace{1.5cm}

                      Wung-Hong Huang$^*$, Kuo-Wei Huang$^{**}$\\
\vspace{1cm}
                       $^*$Department of Physics, National Cheng Kung University\\
                       Tainan, Taiwan\\
\vspace{.5cm}
                     
                       $^{**}$Department of Physics,  National  Sun Yat-Sen University\\
                       Kaohsiung, Taiwan\\

\end{center}
\vspace{1cm}
\begin{center}{\bf  \Large ABSTRACT } \end{center}
The thermodynamics of ideal gas on the noncommutative geometry in the coherent state formalism is investigated.  We first evaluate  the statistical interparticle potential and see that there are residual ``attraction (repulsion) potential" between boson (fermion) in the high temperature limit.  The characters could be traced to the fact that, the particle with mass $m$ in noncommutative thermal geometry with noncommutativity $\theta$  and temperature $T$ will correspond to that in the commutative background with temperature $T(1+kTm\theta)^{-1}$.  Such a  correspondence implies that the ideal gas energy will asymptotically approach to a  finite limiting value as that on commutative geometry at  $T_\theta= (km\theta)^{-1}$.  We also investigate the squeezed coherent states and see that they could have arbitrary mean energy.  The thermal properties of those systems are calculated and compared to each other. We find that the heat capacity of the squeezed coherent states of boson and fermion on the noncommutative geometry  have different values, contrast to that on the commutative geometry.
\vspace{1cm}
\\
\\
\\
\begin{flushleft}
*E-mail:  whhwung@mail.ncku.edu.tw\\
\end{flushleft}


\newpage
\section{Introduction}
The noncommutativity of spacetime which first studied by Snyder [1] is naturally found in  the string/M theories [2,3].  Initially, Connes, Douglas and Schwarz [2]  had shown that the supersymmetric gauge theory on noncommutative torus is naturally related to the compactification of Matrix theory.  It is also known that the dynamics of a D-brane in the presence of  a B-field can, in certain limits, be described by the noncommutative field theories [3]. 

  The noncommutative geometry is currently encoded in the commutator 
$$[ x_\mu, x_\nu] = i \theta_{\mu\nu},\eqno{(1.1)} $$
The value of $\theta_{\mu\nu}$ is an anti-symmetric matrix which determines the fundamental cell discretization of spacetime.   The relation (1) implies the $\star$ operator which is the Moyal product generally defined by
$$\Phi(x) \star \Phi(x) = e^{+{i\over 2} \theta^{\mu\nu} {\partial\over \partial
y^\mu} {\partial\over \partial z^ \nu} } \Phi(y) \Phi(z) |_{y,z\rightarrow x}.  
\eqno{(1.2)} $$
Using above relation the quantum aspects of the noncommutative field theories have been pursued via perturbative analysis over diverse model [4,5].

  Besides the above approach to noncommutative quantum theory there is another method, the coordinate coherent state approach [6,7].  In this approach   the wavefunction of a ``free point particle" becomes
$$ \Psi_{\vec p}(\vec x)= \langle \vec p|\vec x \rangle \sim exp\left(-\theta {\vec p^{~2}\over4}+i \vec p\cdot \vec x\right), \eqno{(1.3)}$$
and many authors had studied the effects of noncommutativity on the terminal phase of black hole evaporation [8-10].   In these study the ``point-like gravitational source" becomes 
$$ \rho(r) = {M\over (4\pi\theta)^{3/2}}exp(-r^2/4\theta).    \eqno{(1.4)}$$
Some interesting results have found  are : there exists a finite maximum temperature that the black hole can reach before cooling down to absolute zero;  there is no curvature singularity at the origin while existence a regular De-Sitter core at short distance. 

    In this short paper we will report our investigations about the thermal property of ideal gas on the noncommutative geometry in the coherent state formalism. 
\section{Thermodynamics of Ideal Gas on Noncommutative Geometry}
\subsection{Statistical Interparticle Potential}
  First, let us  evaluate  the statistical interparticle potential of ideal boson and fermion on the noncommutative geometry.  It is well known that, in comparison with the normal statistical behavior, bosons exhibit a larger tendency of bunching together, i.e., a positive statistical correlation. In contrast, fermions exhibit a negative statistical correlation. Uhlenbeck presented an interesting way of
stating this property by introducing a  ``statistical interparticle potential" and then treating the particles classically [11]. Let us first briefly describe the method [12].  Define the one particle  matrix element of the Boltzmann factor by 
$$F_{ij} = <X_i|e^{-\beta H}|X_j>,\eqno{(2.1)}$$ 
then the matrix element of the Boltzmann factor for a system of two identical particles can be written as 
$$ <X_1,X_2|e^{-\beta H}|X_1,X_2>= F_{11}F_{22}\pm F_{12}F_{21},\eqno{(2.2)}$$ 
where the plus (minus) sign is adopted for the boson (fermion) system. For a translation symmetry system $F_{11}=F_{22}$ and $F_{12}=F_{21}$
and density matrix element becomes [12]
$$<X_1,X_2|\tilde \rho|X_1,X_2>= {1\over V^2}\left[ 1\pm {F_{12}^2\over F_{11}^2}\right], \eqno{(2.3)}$$ 
in which we define the density matrix by $\tilde \rho \equiv {e^{-\beta H}\over \sf {Tr}~e^{-\beta H}}$ [12] and $V$ is the system volume.  The ``statistical interparticle potential" $v$ is defined to be such that the Boltzmann factor exp$(-\beta v)$ is precisely equal to the correlation factor (bracket term) in the above equation, i.e., 
$$v = -kT\ell n \left[1\pm {F_{12}^2\over F_{11}^2}\right]. \eqno{(2.4)}$$

   Now, using the property (1.3) which was found by Smailagic Spallucci [6]
we see that
$$  <X_1|e^{-\beta H}|X_2> = \sum_p <X_1|p><p|e^{-\beta H}|p><p|X_2>\hspace{2.8cm}$$
$$ = {1\over (2\pi)^3} \int d^3k~   e^{-{\hbar^2 \vec k^2\over 2m}\beta +i \vec k\cdot (\vec x_1-\vec x_2)-{\theta \hbar^2 \vec k^2\over 2}} $$
$$= {1\over (2\pi)^3} \int d^3k~   e^{-{\hbar^2 \vec k^2\over 2m}(\beta+m\theta)+i \vec k\cdot (\vec x_1-\vec x_2)}.\eqno{(2.5)}$$
Thus the  ``statistical interparticle potential"  calculated on the noncommutative geometry with inverse temperature $\beta$ is just that calculated on the commutative geometry while with inverse temperature $\beta + m \theta $.

   This means that with the following simple substitution
$$ T ~~~~~ \Rightarrow ~~~~~ f(T) =  {T\over 1+kTm\theta}, \eqno{(2.6)}$$
we could obtain the desired thermal property on the noncommutative background. The ``statistical interparticle potential"  on the noncommutative geometry we found thus becomes 
$$v = - kT ln\left[1\pm exp\left(-{mkT\over\hbar^2(1+kTm\theta)}~r^2\right)\right].\eqno{(2.7)}$$
We plot the above result in figure 1 (with $V\equiv v/kT$) in which the dashed line represents the ``statistical interparticle potential"  on the noncommutative geometry while the solid line represents that on the commutative background.  Boson has negative potential while fermion has positive potential.
\\
\\
\scalebox{1}{\hspace{5cm}\includegraphics{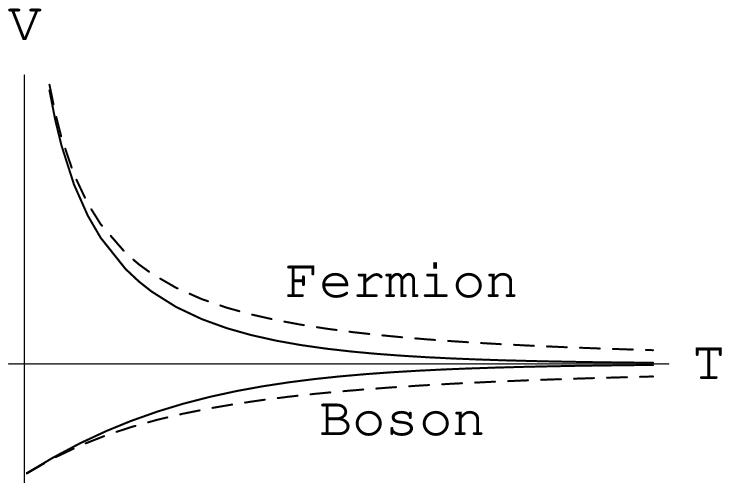}}
\\
\\
{\hspace{1cm} {\it Figure 1:  ``Statistical interparticle potential"  on the noncommutative geometry.  Dashed lines which describe the ideal gas on the noncommutative geometry will asymptotically approach to a finite value.}
\\
\\

Above results show two nontrivial properties:

1.  The space noncommutativity enhances the negative statistical correlation between fermion and enhances the positive statistical correlation between bosons.

2.  There are residual ``attraction potential" between boson and residual ``repulsion potential" between fermion in the high temperature limit.  This is because that, form (2.6) we see that the asymptotic  temperature becomes
$$ {T\over 1+kTm\theta}~~\stackrel{T\rightarrow\infty} {\longrightarrow} ~~{1\over km\theta}\equiv T_\theta. \eqno{(2.8)}$$
Thus the thermal property of the gas on noncommutative background will, in the high temperature, asymptotically approach to the  thermal property of the gas on commutative background with temperature $T_\theta$, which is a finite value.

\subsection{Statistical Mechanics on  Noncommutative Geometry}
In this section we will mention that the simple substitution of eq.(2.6) could be used in many thermal quantities.  First, the key principle of statistical mechanics is that the probability of a point gas having energy $E$ is proportional to $e^{-\beta E}$ [13].  Next, (1.3) tells us that the wavefunction of a ``free point particle" will be smeared to an extended matter by a factor $ exp\left(-\theta {\vec p^{2}\over~4~}\right)$ thus the corresponding probability on the noncommutative geometry will become $e^{-(\beta+m\theta) E}$.  This simple observation implies the simple substituting rule in (2.6). 

Therefore, the energy of ideal fermion or boson could be calculated from the following two relations 
$$ N = \sum_p{1\over z^{-1}e^{(\beta+m\theta)\epsilon}\pm 1}; ~~~U = \sum_p{\epsilon\over z^{-1}e^{(\beta+m\theta)\epsilon}\pm 1},\eqno{(2.9)}$$
where $z$ is  fugacity of the ideal gas, which is related to the chemical potential $\mu$ through the formula $z\equiv exp(\mu/kT)$ [12].  Then, using (2.6) it is easy to see that the Bose-Einstein condensation will occur at 
$$\tilde T_c = T_c \left(1+ {T_c\over T_\theta}\right).\eqno{(2.10)}$$
We see that $\tilde T_c$  is larger then $T_c$ which is the  temperature of Bose-Einstein condensation on commutative background.  The fact that $\tilde T_c > T_c$ could be interpreted from the property that the space noncommutativity enhances the positive statistical correlation between bosons, as shown in figure 1.

   In high temperature limit the energy of  ideal boson (-) or ideal fermion (+)   becomes  
$$ U \approx  {3\over2}Nk{T\over 1+kTm\theta}\pm0.2652{h^3N\over V} {\sqrt{1+kTm\theta}\over\sqrt{2\pi m}\sqrt{kT}}~~$$
$$\stackrel{T\rightarrow\infty} {\approx}{3\over2}NkT_\theta \pm0.2652{h^3N\over V} {1\over\sqrt{2\pi m} \sqrt{kT_\theta}},\hspace{1cm}\eqno{(2.11)}$$
in which the second term corresponds to the effect of residual  ``attraction potential" between boson and residual ``repulsion potential" between fermion in the high temperature limit, as shown in figure 1.   Above result shows a surprising property that, in the coherent state approach, the ideal gas energy at high temperature will asymptotically approach to a finite value, even it is a very large quantity as the space noncommutativity is very small.

   As the appearance of the limiting energy in the above analysis seems something strange it will be interesting to see how it will be modified in the relativistic version.  So, let us consider the property that the particle  is a relativistic one with Einstein relation $ E^2 = p^2+ m^2$,  which may be relevant to the system in the high temperature.  Then, Eq.(2.9) will be replaced by 
$$ N = \sum_p{1\over z^{-1}exp\left({\beta \sqrt{p^2+m^2}}\right)exp\left({\theta p^2/2}\right)\pm 1}. \eqno{(2.12)}$$
$$U = \sum_p{\sqrt{p^2+m^2}\over z^{-1}exp\left({\beta \sqrt{p^2+m^2}}\right)exp\left({\theta p^2/2}\right)\pm 1}.\eqno{(2.13)}$$
Using above relation and  through standard calculation [12] we can find that, for the massless gas in high temperature  we have the following approximations
$$ N/V = {4\pi\over h^3}~\int^\infty_0 dp~p^2 \sum_{\ell =0}^\infty(\pm1)^{\ell}\left(z e^{-\beta\sqrt{p^2+m^2}-\theta p^2/2}\right)^{\ell+1}\hspace{2.4cm}$$
$$\approx {4\pi\over h^3} ~ \left[z \left({\sqrt{\pi }\over \theta^{3/2}\sqrt{2}}-{2\beta\over \theta^2}\right)\pm z^2 \left({\sqrt{\pi }\over 4\theta^{3/2}\sqrt{2}}-{\beta\over \theta^2}\right) \right].\eqno{(2.14)}$$
$$ U/V = {4\pi\over h^3}~\int^\infty_0 dp~p^2 \sqrt{p^2+m^2}\sum_{\ell =0}^\infty(\pm1)^{\ell}\left(z e^{-\beta\sqrt{p^2+m^2}-\theta p^2/2}\right)^{\ell+1}\hspace{0.6cm}$$
$$\approx {4\pi\over h^3} ~ \left[z \left({2\over \theta^2}-{3\beta\sqrt \pi\over \theta^{5/2}\sqrt 2}\right)\pm z^2 \left({1\over 4\theta^2}-{3\beta\sqrt \pi\over 4\theta^{5/2}\sqrt 2}\right) \right].\eqno{(2.15)}$$
Above results imply that  the system energy could be expressed as
$$U = N \left[{ 2\sqrt{2 \pi^{-1}}\over \sqrt\theta} \pm {h^3N\over V} {1-\sqrt 2\over 4\pi^2}~\theta + O(\beta)\right].  \eqno{(2.16)}$$ 
As the relation of relativistic gas energy  on commutative spacetime is $U = 3N kT$ we thus see that the limiting temperature in the relativistic theory becomes
$$ T_\theta = {2\sqrt{2 \pi^{-1}}\over 3 k \sqrt \theta},  \eqno{(2.17)}$$ 
and property of  ``appearing the limiting energy" in the non-relativistic system also shows in the relativistic theory. 

\section{Statistical Mechanics on  Noncommutative Geometry with Squeezed Coherent State}
\subsection{Squeezed Coherent State}
 Let us now turn to consider the squeezed coherent state on the 2D noncommutative space in which the noncommutative coordinates $\hat q_1$ and $\hat q_2$ have the property
$$[\hat q_1,\hat q_2] = i\theta.\eqno{(3.1)}$$
 We first perform the canonical transformation by defining the operators 
$$\hat A= (\hat q_1+ i \hat q_2)\cosh\tau + (\hat q_1- i \hat q_2)\sinh\tau = s \hat q_1+ i s^{-1}~ \hat q_2,$$
$$\hat A^\dag = (\hat q_1+ i \hat q_2)\sinh\tau + (\hat q_1- i \hat q_2)\cosh\tau = s \hat q_1 - i s^{-1}~ \hat q_2,,  \eqno{(3.2)}$$ 
in which the value  $s\equiv e^\tau$ is  the ``squeezing parameter" and the case of $s=1$ is the un-squeezed state considered  in [6,7].  The definition of Eq.(3.2) lead to the relation
$$[\hat A,\hat A^\dag]= 2\theta, \eqno{(3.3)}$$
as that without squeezing. 

As before [6,7]  the squeezed coherent state $ |\alpha_s \rangle$ is defined by  
$$ \hat A|\alpha_s\rangle = \alpha_s|\alpha_s\rangle ~~~~ \langle\alpha_s|\hat A^\dag  = \langle\alpha_s|\alpha_s^*.   \eqno{(3.4)}$$
The  coordinates  $x_1$, $x_2$ which represent the mean position of the particle over the non commutative plane are defined by
 $$x_1 \equiv \langle\alpha_s|\hat q_1|\alpha_s \rangle=  \langle\alpha_s| {\hat A + \hat A^\dag\over 2s} |\alpha_s \rangle={\alpha_s + \alpha_s^*\over2s}.   \eqno{(3.5)}$$
 $$x_2 \equiv \langle\alpha_s|\hat q_2|\alpha_s \rangle=  \langle\alpha_s| {\hat A - \hat A^\dagger\over 2is^{-1}} |\alpha_s \rangle={\alpha_s- \alpha_s^*\over 2is^{-1}}.  \eqno{(3.6)}$$
Note that as $ \hat A|\alpha_s\rangle = \alpha_s|\alpha_s\rangle$ we can define 
$$ \hat A = \alpha_s + \hat \eta, ~~~~~\hat \eta|\alpha_s\rangle = 0,  \eqno{(3.7)}$$
$$ \hat A^\dag = \alpha_s^* + \hat \eta^\dag, ~~~~~\langle \alpha_s|\hat \eta^\dag= 0,  \eqno{(3.8)}$$
in which  $[\eta,\eta^\dag]= 2\theta$.  Thus, defining $\triangle \hat q_1 = \hat q_1 - x_1$ and $\triangle \hat q_2 = \hat q_1 - x_2$  we see that
$$ \langle\alpha_s|(\triangle \hat q_1)^2 |\alpha_s\rangle =  s^{-2}\theta,  \eqno{(3.9)}$$
$$ \langle\alpha_s|(\triangle \hat q_2)^2 |\alpha_s\rangle =  s^2\theta,  \eqno{(3.10)}$$
Thus  the uncertainty on the coordinate  $x_1$ is different from that on the coordinate  $x_2$.  This is what we means the squeezed coherent state.  Note that the squeezed coherent state used in the condensed matter or quantum optics is that on the coordinate  $x$ and ``coordinate'' $p$. 

The noncommutative version of the plane wave operator is defined by $\exp i\left(\vec p\cdot   \vec {\hat q}\right)$ where $\vec p=\left(p_1\, p_2\right)$ is a real two-component vector. After using the Baker-Campbell-Hausdorff  formula the mean value becomes
 $$\langle \alpha_s| e^{ i p_1\hat q_1 +i p_2 \hat q_2}|\alpha_s \rangle =
  \langle \alpha_s|e^{i (p_+ c_+ +p_- c_-)\hat A^\dag +i (p_+ c_- +p_- c_+)\hat A}|\alpha_s \rangle \hspace{5cm}$$
$$\hspace{3.3cm}=
  \langle \alpha_s|e^{i (p_+ c_+ +p_- c_-)\hat A^\dag }e^{i (p_+ c_- +p_- c_+)\hat A}e^{ (p_+ c_+ +p_- c_-)(p_+ c_- +p_- c_+)[\hat A^\dag,\hat A]/2}|\alpha_s\rangle $$
$$\hspace{2.7cm}=
  \langle \alpha_s|e^{i (p_+ c_+ +p_- c_-)\alpha_s^* }e^{i (p_+ c_- +p_- c_+)\alpha_s}e^{- (p_+ c_+ +p_- c_-)(p_+ c_- +p_- c_+)\theta}|\alpha_s\rangle $$
$$= e^{\left(ip_1x_1+ip_2 x_2 -{\theta\over4} ( s^{-2}p_1^2+ s^2p_2^2)\right)}.\hspace{3.8cm}\eqno{(3.11)}$$
In the above calculation we have defined 
$$ p_+\equiv { p_1 + i\, p_2\over 2},~~~~p_-\equiv { p_1 - i\, p_2\over2},\eqno{(3.12)}$$
$$c_+\equiv { s^{-1}+s\over 2},~~~~c_-\equiv { s^{-1}-s\over 2},\eqno{(3.13)}$$
According to the [6,7] we can interpret (3.11)  as the wave function of a ''free point particle'' on the non-commutative plane:
$$ \Psi_{\vec p}(\vec x) =\langle\vec p|\vec x\rangle =exp\left(-\theta {\left(s^{-2} p_1^2 + s^2 p_2^2 \right)\over4}+i \vec p\cdot \vec x\right), \eqno{(3.14)}$$
which is the deformed version of (1.3) 

After the standard calculation [12] we find that 
$$ N = {V\over h^2}~\int^\infty_{-\infty} dp_1dp_2 \sum_{\ell =0}^\infty(\pm1)^{\ell}\left(z e^{-\beta{(p_1^2+ p_2^2)/2m}- \theta(s^{-2}p_1^2+ s^2p_2^2)/2}\right)^{\ell+1}\hspace{5.5cm}$$
$$\approx {V\pi\over h^2} ~ \left({2mkT\over 1+ s^{-2}km\theta T}\right)^{1/2}\left({2mkT\over 1+ s^{2}km\theta T}\right)^{1/2}\left[z \pm {z^2\over2 \sqrt2}\right].\hspace{4cm}\eqno{(3.15)}$$
$$ U = {V\over h^2}~\int^\infty_{-\infty} dp_1dp_2 ~{p_1^2+p_2^2\over 2m}\sum_{\ell =0}^\infty(\pm1)^{\ell}\left(z e^{-\beta{(p_1^2+ p_2^2)/2m}- \theta (s^{-2}p_1^2+ s^2p_2^2)/2}\right)^{\ell+1}\hspace{4cm}$$
$$\approx {V\pi\over 4mh^2} ~\left[ \left({2mkT\over 1+ s^{-2}km\theta T}\right)^{3/2}\sqrt{2mkT\over 1+ s^{2}km\theta T}+ \sqrt{2mkT\over 1+ s^{-2}kmT}\left({2mkT\over 1+ s^{2}km\theta T}\right)^{3/2}\right]\left[z \pm {z^2\over2\sqrt2}\right].\eqno{(3.16)}$$
Using above relations we find that the high-temperature energy of the  squeezed coherent state becomes
$$U ={ N\over2} \left[{kT\over 1+s^2~m\theta kT}+{kT\over 1+~m\theta kT/s^2}\right] \left[1\pm {1\over2\sqrt2}{N h^2\over V\pi}{(2mkT)^2\over (1+s^2~m\theta kT)(1+~m\theta kT/s^2)}\right]$$
$$ \approx {NkT_\theta\over 2} \left[s^2 + {1\over s^2}\right]\left[ 1\pm {\theta\over8\sqrt2}{N h^2\over V\pi}\right]+O(\beta) \approx  NkT_\theta \left[s^2 + {1\over s^2}\right]> NkT_\theta, \hspace{1cm}  \eqno{(3.17)}$$ 
in which $T_\theta$ is defined in (2.8).  Above result implies two interesting properties: 
\\

1.  The squeezed coherent state ($s \ne 1$) has a larger mean energy than the un-squeezed state and it could be copiously excited at high temperature. 

2. The 2D squeezed coherent states of boson and fermion on the noncommutative geometry  have different values of heat capacity, contrast to that on the 2D commutative geometry.

3. We will see in next subsection that when consider the contribution of the arbitrary squeezed coherent state the system will asymptotically approach to infinite energy, contrast to that approaches to a limiting energy in the system with un-squeezed coherent state.
\subsection{Thermodynamics of Squeezed Coherent State}
  To consider the system with arbitrary squeezed coherent state we have to calculate the particle number $N$ and grand potential $q$ which is define by  $ \pm ln(1\pm z e^{-\beta \epsilon})$ in commutative theory [12]. 
$$N = \int^\infty_{-\infty} d\tau \left[ {V\over h^2}~\int^\infty_{-\infty} dp_1dp_2 ~{1\over z^{-1}e^{{p_1^2+p_2^2\over 2m}\beta+\theta\left({e^{-2\tau}p_1^2+e^{2\tau}p_2^2\over 2}\right)} \pm1}\right]$$
$$ = {V\over h^2}~{\pi\over \sqrt{\theta\beta\over m}} P^0_{-{1\over2}}\left({{\beta^2\over4m^2}+{\theta^2\over4}\over{\theta\beta\over2m^2}}\right)~Log\left(1 \pm  z\right),\hspace{2cm}\eqno{(3.18)}$$ 

$$q = \pm \int^\infty_{-\infty} d\tau \left[ {V\over h^2} Log\left(1\pm z e^{-{p_1^2+p_2^2\over 2m}\beta-\theta\left({e^{-2\tau}p_1^2+e^{2\tau}p_2^2\over 2}\right)}\right)\right]\hspace{0.5cm}$$
$$=\mp{V\over h^2}~{\pi\over \sqrt{\theta\beta\over m}} P^0_{-{1\over2}}\left({{\beta^2\over4m^2}+{\theta^2\over4}\over{\theta\beta\over2m^2}}\right)~ {\sf Li_2}\left(\mp z\right),\hspace{0.7cm}\eqno{(3.19)}$$
in which $P^0_{-{1\over2}}(x)$ is the associated Legendre function and ${\sf Li_2}(\mp z)$ is the polylogarithm function.  We can now first use (3.18)  to express the fugacity $z$ as the function of particle number $N$. Next, we can use the energy relation $U = -\partial q/\partial \beta$  and (3.19) to find the system energy.  The results are plotted in figure 2. 
\\
\\

\scalebox{1}{\hspace{3cm}\includegraphics{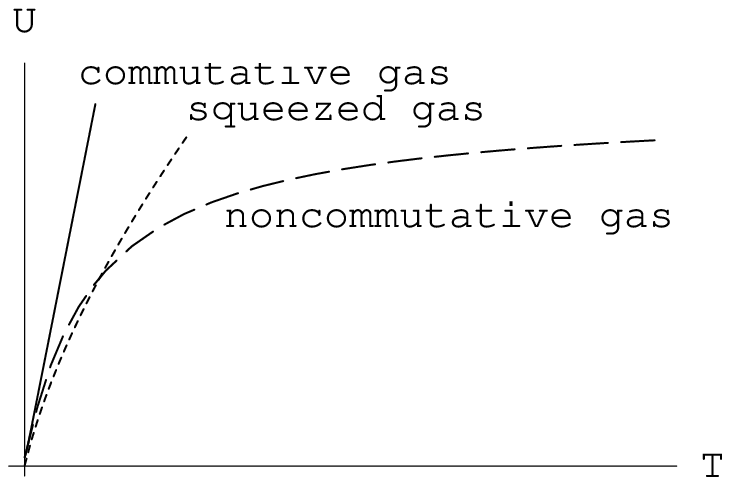}}
\\
\\
{\hspace{1cm} {\it Figure 2 : Energy of the system. Solid line describes the energy of the gas on the commutative space, dashed line describes that of the coherent gas  on the non-commutative space, and dot line describes that of the squeezed coherent gas on the non-commutative space.}
\\
\\
We see that, while the  energy of the coherent gas  on the non-commutative space has a limiting value the squeezed coherent gas could have arbitrary energy.  This property could be seen form the following analytic  formula.   Using the high-temperature expansion 
$$~{\pi\over \sqrt{\theta\beta\over m}} P^0_{-{1\over2}}\left({{\beta^2\over4m^2}+{\theta^2\over4}\over{\theta\beta\over2m^2}}\right)~\approx {2\sqrt\pi \over \theta} + {2\over \theta}\left[2\psi(2)-2\psi(1/2)+ Log\left({\theta m\over 4\beta}\right)\right],\eqno{(3.20)}$$
in which $\psi(x) = {\Gamma'(x)/\Gamma(x)}$ and ${\sf Li_2}(x) \approx x+x^2/4+\cdot\cdot\cdot$ we find that 
$$ U\approx N{\pi kT\over Log(kT)}\left(1\pm{Nk\theta\over 2V Log(kT)}\right),\eqno{(3.21)}$$
which will asymptotically approach to the infinite value.  Above result implies that the heat capacity is
$$C  \approx N{\pi k\over Log(kT)}\left(1- {1\over Log(kT)}\pm{Nk\theta\over 2V Log(kT)}\right).\eqno{(3.22)}$$
Thus, the  heat capacity  of 2D squeezed coherent states of boson and fermion on the noncommutative geometry  have different values, contrast to that on the 2D commutative geometry. We also plot the heat capacity in figure 3.
\\
\\

\scalebox{1}{\hspace{2cm}\includegraphics{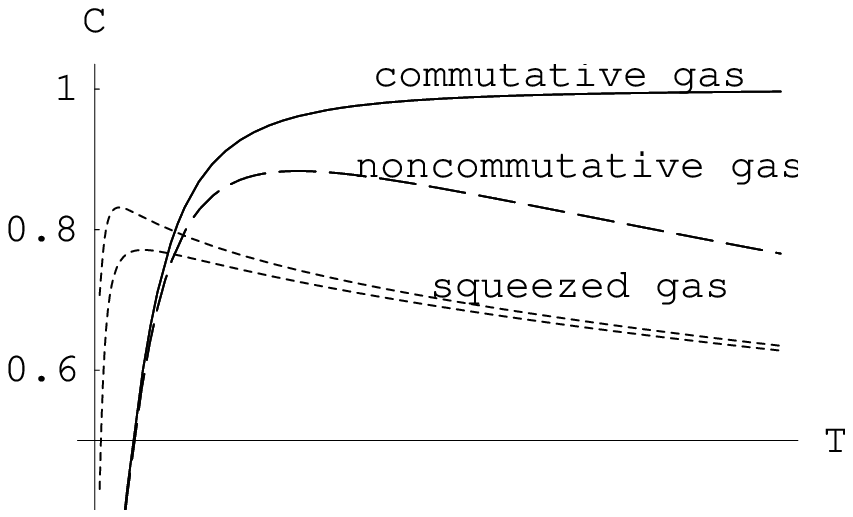}}
\\
\\
{\hspace{1cm} {\it Figure 3: Heat capacity of the system. Solid line describes the heat capacity  of the gas on the commutative space, dashed line describes that of the coherent gas  on the non-commutative space, and dot line (up/below)describes that of the squeezed coherent (fermion/boson) gas on the non-commutative space. }
\\
\\
The figure 3 shows a special property that the heat capacity of the un-squeezed or squeezed coherent state has a peak value at the finite temperature.

\section {Discussion}
In this paper we first see that the particle with mass $m$ in noncommutative thermal geometry with noncommutativity $\theta$  and temperature $T$ will correspond to that in the commutative background with temperature $T(1+kTm\theta)^{-1}$.  Such a  correspondence implies that the ideal gas energy will asymptotically approach to a limiting finite value as that on commutative geometry at  $T_\theta= (km\theta)^{-1}$.  This shows a surprising property that, in the coherent state approach, the ideal gas energy has a limiting energy.  We also discuss the relativistic theory and analyze the squeezed coherent state.  We have found that the heat capacity of the 2D squeezed coherent states of boson and fermion on the noncommutative geometry  have different values, contrast to that on the 2D commutative geometry. Also, when consider the contribution of the arbitrary squeezed coherent state the system will asymptotically approach to the infinite energy, contrast to that approaches to a limiting energy in the system with un-squeezed coherent state.

Finally, Eqs.(3.5), (3.6), (3.9), and (3.10) tell us that, while the un-squeezed and squeezed coherents have the same mean values of coordinates $\hat q_1$ and $\hat  q_2$ they have different values of  coordinate uncertainty.  In the coherent state approach, as the particles we considered are those with the same mean value of coordinate we therefore have to consider both of the un-squeezed and squeezed state.   Such a consideration may dramatically change physical property as we have found.   Along this line, an interesting issue to be investigated is that, as near the final stage of evaporation the backreaction of the black hole on the source will play an important role the black hole source may be deformed into the squeezed coherent state.   Thus the source distribution (1.4) shall be modified to that involved the squeezed state. The detail of the problem remains to be investigated.  
\\
\\
\\
\\
\begin{center} {\bf REFERENCES}\end{center}
\begin{enumerate}
\item H. S. Snyder, ``Quantized Space-Time'', Phys. Rev. 7 (1947) 38;   72 (1947) 68;  Connes, ``Noncommutative Geometry'', Academic. Press, New York, (1994).
\item  A. Connes, M. R. Douglas and A. Schwarz,  ``Noncommutative Geometry and Matrix Theory: Compactification on  Tori'', JHEP 9802 (1998) 003 [hep-th/9711162]. 
\item  N.~Seiberg and E.~Witten, ``String Theory and Noncommutative Geometry,''  JHEP 9909 (1999) 032 [hep-th/9908142].
\item   M. R. Douglas and N. A. Nekrasov, ``Noncommutative Field Theory ,'' Rev. Mod.Phys. 73 (2001) 977 [hep-th/0106048 ]. 
\item   R. J. Szabo, ``Quantum Field Theory on Noncommutative Spaces ,''  Phys.Rept. 378 (2003) 207 [hep-th/0109162].
\item  S. Cho, R. Hinterding, J. Madore, H. Steinacker,``Finite Field Theory on Noncommutative Geometries," Int.J.Mod.Phys. D9 (2000) 161 [hep-th/9903239];  
A. Smailagic and E. Spallucci,``Feynman Path Integral on the Noncommutative Plane, " J.Phys. A36 (2003) L467 [hep-th/0307217 ]; ``UV divergence-free QFT on noncommutative plane,"  J. Phys. A36 (2003) L517 [hep-th/0308193 ]. 
\item  A. Smailagic and E. Spallucci,``Lorentz invariance and unitarity in UV finite NCQFT ,"  J.Phys. A37 (2004) 1 [hep-th/0406174];  A. Gruppuso,``Newton's law in an effective non commutative space-time ," J.Phys. A38 (2005) 2039 [hep-th/0502144 ].
\item   P, Nicolini,`` A model of radiating black hole in noncommutative geometry ," J.Phys. A38 (2005) L631 [hep-th/0507266]; P. Nicolini, A. Smailagic, and E. Spallucci ,``Noncommutative geometry inspired Schwarzschild black hole ," Phys.Lett. B632 (2006) 547 [gr-qc/0510112]; S. Ansoldi, P. Nicolini, A. Smailagic, and E.Spallucci ,``Noncommutative geometry inspired charged black holes ," Phys.Lett. B645 (2007) 261 [gr-qc/0612035].
\item  Y. S. Myung, Y.-W. Kim, and Y.-J. Park ,``Thermodynamics and evaporation of the noncommutative black hole ,"  JHEP 0702 (2007) 012 [gr-qc/0611130]; R. Banerjee, B. R. Majhi, S. Samanta,`` Noncommutative Black Hole Thermodynamics,"  [arXiv:0801.3583]; R. Banerjee, B. R. Majhi, S. K. Modak ,``Area Law in Noncommutative Schwarzschild Black Hole,"  [arXiv:0802.2176].
\item   P, Nicolini,``Noncommutative Black Holes, The Final Appeal To Quantum Gravity: A Review," [arXiv:0807.1939 ].
\item   G. E. Uhlenbeck and L. Gropper, Phys. Rev. 41 (1932) 79 .
\item  R. K. Pathria, Statistical Mechanics (Pergamon, London, 1972).
\item  R. P. Feynman, Statistical Mechanics (1972).
\end{enumerate}
\end{document}